\documentclass{pasj01}
\usepackage[switch,mathlines]{lineno}
\Received{$\langle$11-Jan-2024$\rangle$}
\Accepted{$\langle$7-May-2024$\rangle$}
\Published{$\langle$publication date$\rangle$}

\begin{document}

\title{A Comprehensive Study of an Oscillating Eclipsing Algol: Y\,Cam}
\author{Eda \textsc{\c{C}elik}\altaffilmark{1}\thanks{E-mail: celikeda7575@gmail.com, filizkahraman01@gmail.com }
and Filiz \textsc{Kahraman Ali\c{c}avu\c{s}}\altaffilmark{2,3}}
\altaffiltext{1}{\c{C}anakkale Onsekiz Mart University, School of Graduate Studies, Physics Department, TR-17100, Çanakkale, Turkey}
\altaffiltext{2}{\c{C}anakkale Onsekiz Mart University, Faculty of Science, Physics Department, TR-17100, Çanakkale, Turkey}
\altaffiltext{3}{\c{C}anakkale Onsekiz Mart University, Astrophysics Research Center and Ulup{\i}nar Observatory, TR-17100, Çanakkale, Turkey}



\maketitle

\begin{abstract}

Y\,Cam is classified as one of the oscillating Eclipsing Algol (oEA) systems, which feature a $\delta$\ Scuti-type pulsating component alongside mass transfer phenomena. oEA systems are invaluable for probing the evolutionary processes and internal structures of binary components offering insights through their binary variations and oscillating. In this study, we conducted a comprehensive investigation of Y\,Cam utilizing high-quality photometric TESS data, and high-resolution ELODIE spectra. Through our analysis, we examined the radial velocity variation, performed binary modeling, and calculated the effective temperature values of binary components. The fundamental stellar parameters, such as mass and radius, were determined with an accuracy of $\sim$$2-6$\%. Furthermore, we examined the orbital period variation to assess the amount of mass transfer using the available minima times of the system and three new minima times obtained from TESS light curves. Analyzing the pulsation structure of the system with the TESS data revealed the dominant pulsation period and amplitude of the pulsating component to be 0.066\,d and 4.65\,mmag, respectively. Notably, we observed frequency modulations with the orbital period's frequency, along with variations in the amplitude of the highest amplitude frequency across different orbital phases. Remarkably, the amplitude reaches its peak at phases 0.5 and 1. These findings indicate a candidate of a tidally tilted pulsator. Consequently, we investigated the evolutionary status of the binary components using MESA binary evolution models, determining the age of the system to be 3.28\,$\pm$\,0.09\,Gyr.


\end{abstract}
\section{Introduction}

The significance of the eclipsing binary (EB) systems has been widely known for decades. These systems provide a unique opportunity to determine the fundamental stellar parameters (such as mass $M$, radius $R$) with 1$\%$\ accuracy \citep{2010A&ARv..18...67T, 2013A&A...557A.119S}. This sensitivity is essential to truly examine the stellar structure and evolution. To achieve the desired accuracy the photometric light curve and radial velocity variations of an EB system should be analyzed simultaneously. 

There are various types of EB systems classified based on the Roche geometry into three primary groups: detached, semi-detached, and contact EB systems \citep{1959cbs..book.....K}. Among these, semi-detached binary systems are more compelling systems as they demonstrate mass transfer between the binary components. Algol-type binaries represent a unique subset within the semi-detached EB systems. Typically, Algols consist of one more luminous and massive B to F type star alongside a less luminous and massive evolved giant or subgiant companion \citep{1959cbs..book.....K}. The evolution of the Algol-type systems is notable given that currently, the less massive components appear more evolved compared to the other massive binary component. Additionally, these systems undergo mass transfer between the binary components, and this profoundly affects the evolution of the systems. Therefore, a thorough examination of Algol-type systems is essential to gaining deeper insights into the binary evolution. 

Among the Algol-type systems, there are some remarkable ones that exhibit pulsations alongside binary variations in their light curve. Pulsating stars undergo oscillations induced mostly by the $\kappa$ mechanism triggered by the partial ionization zone of elements like hydrogen, helium, and iron \citep{2010aste.book.....A}. These systems offer a unique opportunity to probe the internal stellar structure. Stellar oscillations enable investigations of the stellar rotation, convective core to overshoot mixing, and internal rotation by using stellar oscillations \citep{2023ApJ...948...16K, 2023arXiv230516721Z, 2017MNRAS.465.2294O}. Pulsating stars display diverse evolutionary statuses and are expected to show oscillations within a specific area in the Hertzsprung-Russell diagram that is known as their instability strip \citep{2010aste.book.....A}. Some of them are commonly known pulsating systems such as $\delta$\ Scuti ($\delta$\ Sct), $\beta$\ Cepheids ($\beta$\ Cep), and $\gamma$ Doradus ($\gamma$ Dor) stars. As mentioned before, these pulsators can be observed in EB systems as their components \citep{2021Galax...9...28L,2021Univ....7..369S}. Among these pulsating stars, $\delta$\ Sct pulsators are frequently detected in EB systems due to their relatively shorter pulsation periods and pulsation amplitude. $\delta$\ Sct variables are A- to F-type systems spanning from dwarf to giant luminosity classes. Typically, they exhibit oscillations with pulsation periods ranging from 18 minutes to 8 hours manifesting in low-order radial and non-radial pressure and gravity modes with pulsation amplitude below 0.1 in V-band \citep{2010aste.book.....A}. Notably, many $\delta$\ Sct variables have been identified within semi-detached Algol type eclipsing binary systems in addition to detached binaries \citep{2017MNRAS.470..915K,2017MNRAS.465.1181L}.

The main-sequence component of Algols has the potential to exhibit oscillations as they can fall within the classical pulsating instability strip. Pulsating stars in Algol-type semi-detached binary systems are termed oscillating Eclipsing Algol (oEA). These oEA stars are characterized as "mass-accreting main sequence A/F-type components in semi-detached Algols that show $\delta$ Sct-like pulsation" \citep{2002ASPC..259...96M,2004A&A...419.1015M}. Numerous oEA systems have been identified and investigated in the literature, with their count increasing due to recent discoveries \citep{2017MNRAS.470..915K,2022ApJS..263...34C, 2022ApJS..259...50S, 2022Galax..10...97M, 2022RAA....22h5003K, 2023MNRAS.524..619K}. A notable characteristic of oEA stars is the ongoing mass transfer from the inner Lagrange L1 point leading to the accretion of transferred mass onto the atmosphere of the pulsating component. Such mass transfer events have the potential to significantly impact the pulsation structure of oscillating binary components. Recent research has demonstrated that the effect of mass transfer can be discerned in the pulsation frequencies \citep{2024arXiv240305627W}. The impact of mass transfer on pulsations in oEA systems was anticipated years ago \citep{2002ASPC..259...96M}. In these systems, the transferred mass from the evolved star to the pulsating component interacts with the atmospheres of pulsating stars, exchanging angular momentum and altering surface differential rotation which can lead to difference in oscillations \citep{2008ApJ...679.1499L, 2018MNRAS.475.4745M}.\,\citet{2018MNRAS.475.4745M} provided the first demonstration of how mass transfer affects the pulsation amplitude and frequencies by investigating the well-known oEA system RZ\,Cas. In the literature, several well-known examples of oEA systems have been extensively studied by researchers (e.g., \cite{2022MNRAS.514..622M,2020AJ....160..247P}). RZ\,Cas is one of those systems that is an active and luminous Algol-type system that has undergone detailed studies both photometrically and spectroscopically. Through precise high-resolution observations, \citet{2020A&A...644A.121L} elucidated the orbital period and radial velocity (RV) variations of RZ\,Cas by using cool spot models on the surface of the secondary component, particularly facing the first (L1) and second (L2) Lagrangian points. Additionally, they explored the correlation between spot sizes and mass transfer activity, revealing that the mass transfer activity is modulated by variable-depth Willson depression on the L1 spot. The case of RZ\,cas shows us the importance of thoroughly investigating oEA systems. Research on such systems is essential for gaining a deeper understanding of the influence of mass transfer on pulsation behavior and binary evolution. Therefore, in this study, we focus on an oEA system of Y\,Cam.


Y\,Cam (V\,=\,10$^{m}$.60, A9IV\,+\,K1IV \cite{2001A&A...366..178R}) is classified as an Algol-type EB system \citep{1903AN....162..205C}. The pulsational behavior of this system was initially detected by \citet{1973IBVS..823....1B} who determined the highest pulsation amplitude to be 0$^{m}$.04 with a period of 0.063\,day. Subsequently, analysis of BV band photometric observations by \citet{1974A&A....34...89B} confirmed the $\delta$\ Sct-type oscillations of the primary component, unveiling its multi-periodic pulsating behavior. \citet{2002A&A...391..213K} further corroborated this finding, reporting four pulsation frequencies. Estimation of the fundamental stellar parameters of Y\,Cam was carried out by \citet{2010MNRAS.408.2149R} through binary modeling using data obtained from Str\"{o}grem $uvby$, Jonson V, and Crawford H$_{\beta}$ filters. Subsequently, physical parameters were determined by analyzing the double-lined radial velocity ($v$$_{r}$) curves in conjunction with the photometric data \citep{2015AJ....150..131H}. This study suggested that both binary components have undergone mass exchange with the primary component experiencing an evolution path distinct from that of single $\delta$\ Sct-type pulsators.

Y\,Cam has been the subject of some observational and theoretical investigations over the years, encompassing photometric, spectroscopic, and astrometric studies (e.g. \cite{2015AJ....150..131H,2002A&A...391..213K,2010MNRAS.408.2149R}). However, none of these studies were conducted with the aid of high-quality photometric data. Therefore, in this study, we present an analysis utilizing high-quality space-based data in conjunction with new spectroscopic analysis and radial velocity measurements from the literature. The paper is structured as follows. Sect.\,2 introduces the high-quality photometric and new spectroscopic data. The analysis of the radial velocity measurements is presented in Sect.\,3. In Sect.\,4 determination of the atmospheric parameters is explained. The results of the orbital period analysis are given in Sect.\,5. The binary modeling and the determination of the fundamental stellar parameters are presented in Sect.\,6. The frequency analysis is given in Sect.\,7. In Sec.\,8, the evolution modeling of the system is given. Discussion and conclusions are presented in Sect.\,9.

\section{Observational Data}

In this study, we used data from Transiting Exoplanet Survey Satellite (TESS) for the binary modeling and frequency analysis of Y\,Cam. TESS is a space telescope launched to discover new exoplanets in bright and nearby stars \citep{2015JATIS...1a4003R}. TESS observes the sky continuously, dividing it into sectors, each lasting approximately 27 days. TESS collects data at both short (SC) and long cadences (LC). SC data are acquired every 120s, while LC data are obtained at intervals of 200s, 600s, and 1800s. TESS provides two types of fluxes: simple aperture photometry (SAP) and pre-search aperture photometry (PDCSAP) \citep{2015JATIS...1a4003R}. The TESS data used in the study were obtained from Milkuski Archive for Space Telescopes (MAST)\footnote{https://mast.stsci.edu/} from sectors 40, 47, and 53. We opted to use SAP fluxes from SC data due to their lower uncertainties, and the Nyquist frequency of SC data, approximately  360\,d$^{-1}$, is suitable for detecting $\delta$\, Sct-type pulsations. Each flux value was converted into magnitude utilizing the same method used by \citet{2022RAA....22h5003K}.

We also conducted searches in spectral databases and found four public ELODIE spectra of Y\,Cam. However, due to the low signal-to-noise ratio (SNR) of two of those spectra, only two were suitable for inclusion in the analysis. ELODIE was a fiber-fed \'{e}chelle spectrograph mounted on the 1.93-m telescope at the Observatoire de Haute–Provence (OHP), specifically designed for obtaining high-accuracy radial velocity measurements. The original spectra cover a wavelength range of 389\,$-$\,618\,nm across 67 orders and boast a resolving power of R\,=\,42000 \citep{2004PASP..116..693M}. The ELODIE archive\footnote{http://atlas.obs-hp.fr/elodie/} provides spectra in two different formats: \emph{S2D} and \emph{spec}. The \emph{S2D} format records spectra with a step of 0.05, comprising 67x1024 pixels. The \emph{spec} format covers a wavelength range of 400\,$–$\,680 nm, accessible as instrumental and continuum normalized flux \citep{2004PASP..116..693M}. For our study, we utilized two ELODIE spectra observed on 11/01/2003 with SNR values of around 25. We employed instrumental flux in \emph{S2D} format and normalized spectra using HANDY\footnote{https://github.com/RozanskiT/HANDY} program. Given the low SNR, spectroscopic data was solely used for radial velocity ($v$$_{r}$) measurement and for determining the effective temperature ($T_{\rm eff}$) values of binary components using the H$\alpha$ line.

\begin{center}
    \centering
    {\scriptsize
    \begin{table}
    \caption{Results of the radial velocity curve analysis of Y\,Cam.}
    \label{tab:table_1}
    \begin{tabular}{p{3.5cm}p{3.5cm}}
        \hline\noalign{\vskip3pt}
        Parameter               &  Value  \\
        \hline\noalign{\vskip3pt}
        $P_{orb}$ (d)		    &$3.305735^a$\\
        $T_0$ (HJD)	            &2452502.242 $\pm$ 0.013\\
        $e$			            &$0.0^a$\\
        $\omega$ (deg)          &$90.0^a$ \\
        $\gamma$ (km/s)         &-7.07 $\pm$ 0.92\\
        $q = M_2/M_1$           &0.237 $\pm$ 0.010\\
        $a_{1}$sin$i$ ($R_\odot$) &2.037 $\pm$ 0.134\\
        $a_{2}$sin$i$ ($R_\odot$) &10.270 $\pm$ 0.252\\
        $a$sin$i$ ($R_\odot$)     &12.702 $\pm$ 0.267\\
        $K_1$ (km/s)            &37.23 $\pm$ 1.33\\
        $K_2$ (km/s)            &157.24 $\pm$ 3.86\\
        \hline\noalign{\vskip3pt}
    \end{tabular}

    \begin{tabnote}
    {\textbf{Notes:} Subscript $1$ and $2$ represent the primary and secondary components, respectively. Superscript $a$ represents the fixed parameters and $i$ shows the inclination.}
    \end{tabnote}
    \label{table:test1}
\end{table}
}
\end{center}

\begin{figure}
 \begin{center}
  \includegraphics[width=80mm]{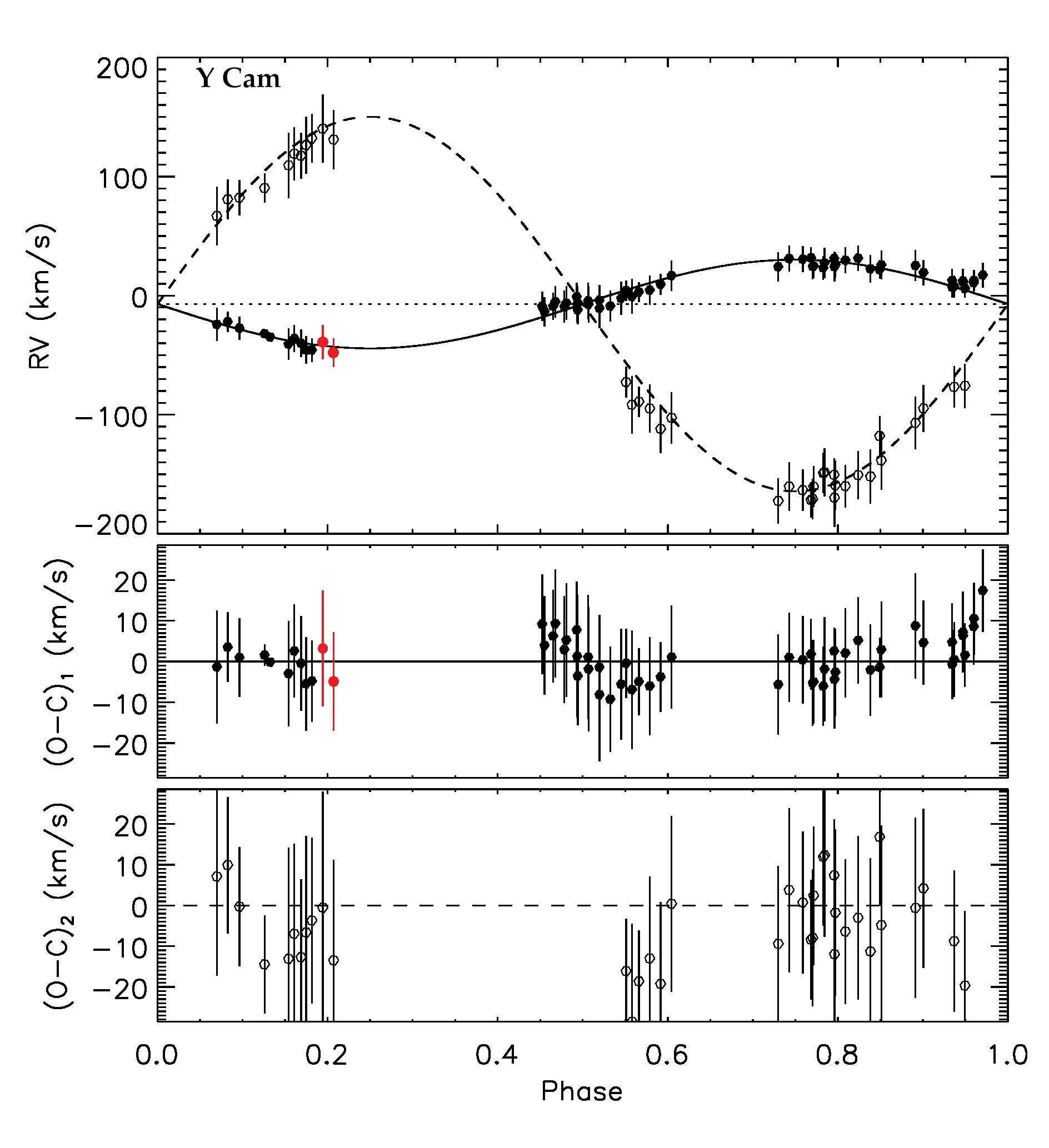}
 \end{center}
 \caption{Upper panel: Radial velocity curve fits to the literature data (black dots, \cite{2015AJ....150..131H}) and our measurements (red dots) of $v$$_{r}$. Middle panel: Residuals (O-C) from the primary (more luminous star) $v$$_{r}$ values. Lower panels: Residuals (O-C) from the secondary (less luminous star) $v$$_{r}$ values.}\label{fig:figure1}
\end{figure}

\section{Radial Velocity Analysis}

To update the orbital parameters of Y\,Cam a radial velocity ($v$$_{r}$) analysis was carried out. The FXCOR task of IRAF/NOAO package \citep{1986SPIE..627..733T} was utilized to calculate the $v$$_{r}$ values of the ELODIE spectra. This task facilitates the measurement of $v$$_{r}$ values by comparing a template spectrum with the observed spectrum using the cross-correlation method. Initially, the heliocentric offset was determined using the RVCOR task in IRAF/NOAO package and the offset was subsequently corrected using the DOPCOR task for both spectra. During the $v$$_{r}$ measurement process, a synthetic spectrum with the parameters of $T_{\rm eff}$\,=\,8000\,K, surface gravity $\log g$\,=\,4.0, and metallicity [m/H]\,=\,0.0 was employed as a template. The rotational velocity of the synthetic spectrum was set to \emph{v}sin\emph{i}\,=\,50 km/s taking into account the values given in the study of \citet{2015AJ....150..131H}. The ATLAS9 model atmospheres \citep{1993KurCD..13.....K} and SPECTRUM code\,\footnote{http://www.appstate.edu/~grayro/spectrum/spectrum.html} \citep{1994AJ....107..742G} were used to generate the synthetic spectrum. Due to the very low SNR, $v$$_{r}$ values could only be measured for the primary components. The $v$$_{r}$ values were determined as -46.8\,$\pm$\,3.5 km/s at HJD 52650.912454 and -34.2\,$\pm$\,4.6 km/s at HJD 52650.93845 from the first and second spectra, respectively.

 The measured $v$$_{r}$ values were analyzed by incorporating them with the literature $v$$_{r}$ values \citep{2015AJ....150..131H}. The \emph{rvfit} code \citep{2015PASP..127..567I} was used in the analysis. During the analysis, orbital period ($P_{orb}$) and time of phase zero ($T_0$) were taken to be 3.305740\,(2)\,d and 2452502.202\,(2) HJD as input, respectively \citep{2004AcA....54..207K}. The $P_{orb}$, orbital eccentricity ($e$), and the argument of the periastron ($\omega$) are held fixed parameters during the analysis, as $e$ was previously determined to be 0 \citep{2015AJ....150..131H}. Meanwhile, $T_0$, the systemic radial velocity $\gamma$, and the amplitude of radial velocities variations ($K_{1}$, $K_{2}$) of the primary and secondary stars were taken as free parameters in the analysis. The results of the analysis are presented in Table~\ref{tab:table_1} and the theoretical $v$$_{r}$ fit to the observed $v$$_{r}$ values is shown in Fig.\,~\ref{fig:figure1}. In the figure, black-filled and open dots represent the literature $v$$_{r}$ measurements of primary and secondary components, respectively, while red-filled dots represent our $v$$_{r}$ measurements for the primary component. According to the result of the analysis, the parameters are consistent with the literature within the range of uncertainties. Since the program we used for the $v$$_{r}$ analysis does not take into account the Rositter-McLaughlin effect, the result of an $v$$_{r}$ analysis could show slight differences from the literature.

\begin{figure*}
 \begin{center}
  \includegraphics[width=150mm]{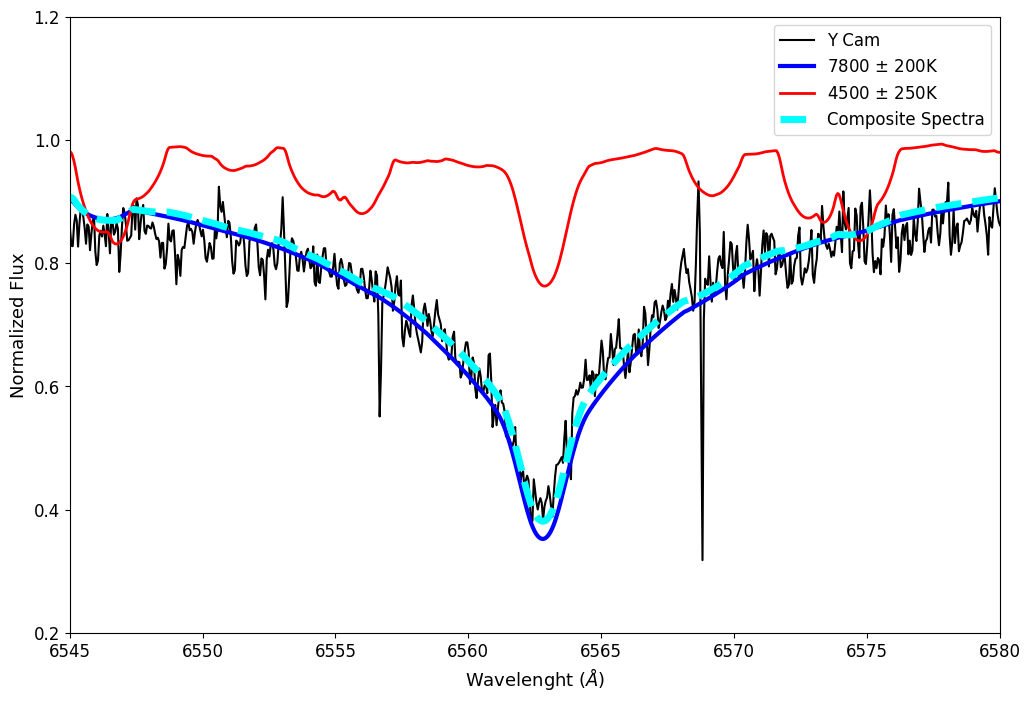}
 \end{center}
 \caption{The theoretical composite H$\alpha$ line (dark blue solid line) fit to the observed composite H$\alpha$ line (black line) of Y\,Cam. Spectra of the primary and secondary components are shown with light blue and red solid lines, respectively.}\label{fig:figure2}
\end{figure*}

\section{Determination of $T_{\rm eff}$ Parameters}

Determination of accurate $T_{\rm eff}$ values of binary components is essential to calculate precise fundamental stellar parameters, especially for estimation of the bolometric ($M_{bol}$) and absolute magnitudes ($M_{v}$). Therefore, we derived $T_{\rm eff}$ parameters by creating composite spectra considering the flux ratio of the primary and secondary components. In this analysis, we utilized the hydrogen alpha (H$_{\alpha}$) line, because the SNR level is higher in this part of spectra. So that will allow us to estimate more sensitive results. To determine initial $T_{\rm eff}$ parameters for the analysis, initially, we determined the $T_{\rm eff}$ of the system by using Johnson UBV and 2MASS photometry. We utilized the method given by \citet{2000AJ....120.1072S} and by \citet{2006A&A...450..735M} for calculating $T_{\rm eff}$ from UBV and 2MASS photometry, respectively.
The system's $B$, $V$, and $K$ magnitudes are 10.93, 10.60, and 9.47\,mag, respectively (\cite{2000A&A...355L..27H,2003yCat.2246....0C}). Before the calculation of $T_{\rm eff}$ from color indexes, first, we determined interstellar reddening of $E(B-V)$\,=\,$0^m$.078\,$\pm$\,0.002 by using the dust map \citep{2019ApJ...887...93G}. Then, with the help of the equation given in the study of \citet{1996A&AS..117..227A}, the $E(V-K)$ value was found to be $1^m$.109\,$\pm$\,0.003. Consequently, the $T_{\rm eff}$ of the system was calculated as 7427\,$\pm$\,275 K and 6820\,$\pm$\,65 K from the $(B - V)_{\rm 0}$ and $(V - K)_{\rm 0}$ indexes, respectively. Since the primary component has over 90\% flux contribution in total according to the study of \citet{2015AJ....150..131H}, these calculated $T_{\rm eff}$ values were considered an initial $T_{\rm eff}$ value for the primary component. 

After obtaining the initial $T_{\rm eff}$ parameter from color indexes and also considering 7396\,K $T_{\rm eff}$ value from the TESS Input Catalog (TIC, \cite{2019AJ....158..138S}), we generated synthetic spectra in the range of 6800\,K\,$-$\,8200\,K in 100\,K step for the primary and in the range of 4250K\,$-$\,4750K in 250\,K step for the secondary component. For $T_{\rm eff}$ range of the secondary component, we took into account the $T_{\rm eff}$ values given in the literature binary modelings for the secondary star \citep{2015AJ....150..131H,2002A&A...391..213K,2010MNRAS.408.2149R}. 

The ATLAS9 model atmospheres \citep{1993KurCD..13.....K} and SPECTRUM\,\footnote{http://www.appstate.edu/~grayro/spectrum/spectrum.html} code were used \citep{1994AJ....107..742G} to generate synthetic spectra. During the generation, the projected rotational velocity ($v\sin i$) value was set to be 50 km/s based on the study of \citet{2015AJ....150..131H} using AVSINI code \citep{1994AJ....107..742G}. The $\log g$ was fitted to 4.0 because the $\log g$ value does not affect the profile of hydrogen Balmer lines at $T_{\rm eff}$ below 8000K \citep{2014dapb.book...97C}. Also, we set the metallicity to [m/H] = 0.0 as the hydrogen lines are not deeply affected by metallicity. We determined $T_{\rm eff}$ by combining synthetic spectra according to the flux contribution of binary components. The flux contributions were taken as 0.93 for primary and 0.07 for secondary according to the V band light curve analysis in the study of \citet{2015AJ....150..131H}. As a result, the $T_{\rm eff}$ of the primary component was determined as 7800\,$\pm$\,200\,K and the $T_{\rm eff}$ of the secondary component as 4500\,$\pm$\,250\,K. The uncertainties in the $T_{\rm eff}$ corresponds to 1$\sigma$. The consistency between the theoretical and observed composite spectrum is given in Fig.~\ref{fig:figure2}.

\begin{figure}
 \begin{center}
  \includegraphics[width=75mm]{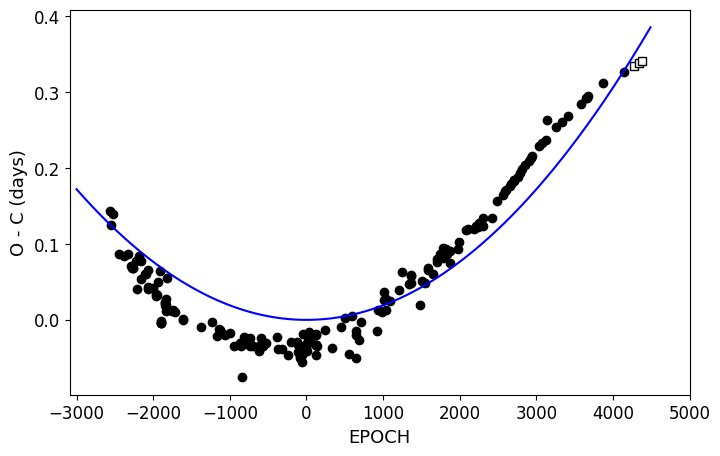}
   \includegraphics[width=75mm]{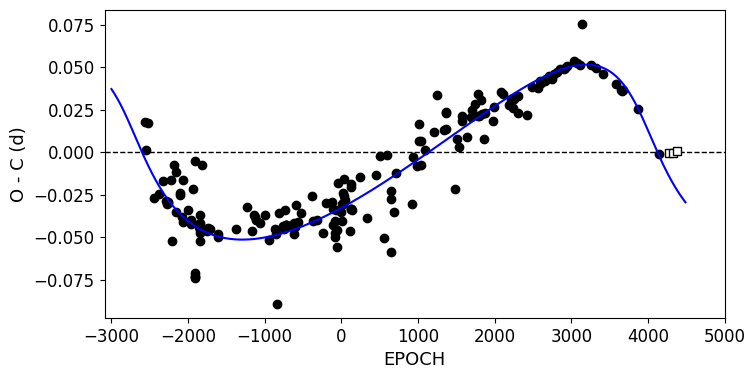}
  \includegraphics[width=75mm]{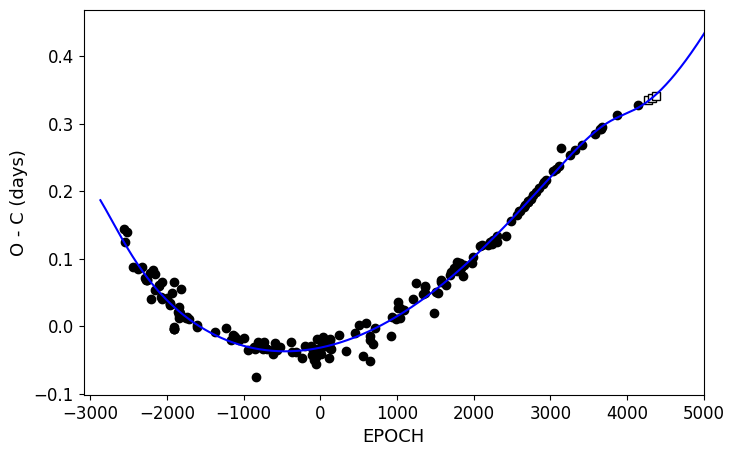}
 \end{center}
 \caption{$O\,-\,C$ variation (dots) and theoretical fit (solid blue line). The upper panel shows the parabolic fit, while the middle panel displays residuals from the parabolic fit and the LITE model fit to these residuals. The lower panel exhibits the combination of the parabolic variation and LITE model fit.  The black squares represent the minimum times estimated in this study.}\label{fig:figure3}
\end{figure}

\section{Orbital Period Analysis}

The orbital period of a binary system may undergo slight variations due to several factors, including the influence of a third celestial body, magnetic activity, mass transfer between the components, and potential mass loss from the system. Y\,Cam is notably among the systems that predominantly exhibit an augmentation in the orbital period \citep{2013ARep...57..517K}. This increase typically signifies a process of mass transfer occurring between the binary components. Therefore, in this section, we examine the orbital period variation of Y\,Cam to understand the nature of the orbital period change of the system. For this purpose, we collected all available literature minima times from the $O\,-\,C$ Gateway\,\footnote{http://var2.astro.cz/ocgate/}. 
Additionally, we conducted measurements of three new minima times utilizing TESS sectors of 40\,-\,47\,-\,53. Each sector contributed one extracted from the middle portion of the observed data. The calculation of new minima times was performed following the method outlined by \citet{1956BAN....12..327K}. Since TESS observation has been carried out last seven years, all minima times correspond to almost the same region on the "Observed minus Calculated ($O\,-\,C$)" diagram which shows the orbital period changes.

Our analysis focused on data collected after $Epoch (E)$ -\,2562 as these datasets exhibited less scattering and higher reliability. A total of 94 visual, 18 photographic, 26 photoelectric, and 46 CCD minima times including TESS minima were incorporated into the study. We assigned data weights of 1 and 4 for visual and photographic minima times, respectively, while assigning a weight of 10 to photoelectric and CCD minima times.

The $O\,-\,C$ diagram of Y\,Cam displays prominent upward parabolic variations, which may be attributed to mass transfer. However, upon closer examination, it can be seen that there are additional variations present alongside the parabolic changes. Therefore, we initially applied a second-order polynomial fit to $O\,-\,C$ data by using following formula:
\\

$(O-C)=191.534\,\times \,10^{-10} (5) E^2 + 3.305654 (5)\,\times\,E \newline + 2445259.391(9)$
\\

Here, $E$ is the number of orbital cycles, which can be computed based on a given reference minima time. Subsequently, we determined a rate of increase in the orbital period of 0.366 $s\,yr^{-1}$. The quadratic term was found as $Q = 191.534\,\times \,10^{-10}$ d, and employing a non-conservative mass-transfer assumption \citep{2014MNRAS.441.1166E} the amount of mass transfer between components was calculated as $3.86\,\times\,10^{-8} M_\odot$. The parabolic fit to the $O\,-\,C$ data is illustrated in the upper panel of Fig.\,\ref{fig:figure3}.




After modeling the parabolic variation, we observed that the residual data displays a cyclic variation. This variation was further examined with the possible cause being attributed to the LIght Time Effect (LITE) caused by a third body orbiting around the center of mass. The observed cyclic variation could potentially be the result of a magnetic activity. However, it is noteworthy that in such cases, the cyclic variation tends not to exhibit a smooth sinusoidal shape. The 'Applegate Mechanism' explains the irregular cyclic variation caused by magnetic activity \citep{1992ApJ...385..621A}. Hence, the obtained cyclic variation was attributed to be caused by a third body. The association between the LITE and $O\,-\,C$ was initially established by \citet{1959AJ.....64..149I}. Using the equations given in this study, we modeled the residual cyclic variations, and the period, minimum mass of the third body were determined to be 60.3\,$\pm$\,1.1 years, and 1.74\,$\pm$\,0.04\,$M_\odot$, respectively. The resulting analysis fit to these residuals is shown in the middle panel of Fig.\,\ref{fig:figure3}. The combined fit of parabolic and cyclic variation to the whole $O\,-\,C$ diagram is also given in the lower panel of Fig.\,\ref{fig:figure3}. The parameters obtained from the orbital period analysis are presented in Table~\ref{tab:table_2}.


\begin{center}
    \centering
    {\scriptsize
    \begin{table}
    \caption{Results of the orbital period analysis. Subscript 3 represents the third celestial body.}
    \label{tab:table_2}
    \begin{tabular}{lc}
    \hline\noalign{\vskip3pt}
    Parameter     &Value \\
    \hline\noalign{\vskip3pt}
    $T_0$ (HJD)   &2445259.391\,$\pm$\,0.009  \\
    $P$ (day)     &3.305654\,$\pm$\,0.000005    \\
    $Q$ (day) ($10^{-10}$)    &191.534\,$\pm$\,0.005    \\
    $dP/dt$ ($syr^{-1}$)    &0.366\,$\pm$\,0.009 \\
    $dM/dt$ ($M_\odot yr^{-1}$) ($10^{-8}$)    &3.86\,$\pm$\,0.08 \\
    $P_3$ (yr)  &60.3\,$\pm$\,1.1   \\
    $T_3$ (HJD) &2414195\,$\pm$\,673 \\
    $A_3$ (day)   &0.049\,$\pm$\,0.002  \\
    $\omega_3$(deg)   &164\,$\pm$\,5 \\
    $e_3$     &0.56\,$\pm$\,0.09  \\
    $f(m_3)$ ($M_\odot$)    &0.288\,$\pm$\,0.001  \\
    $m_3$($M_\odot$) ($i = 90)$ &1.74\,$\pm$\,0.04 \\
    \hline\noalign{\vskip3pt}
    \end{tabular}
    \label{table:test2}
\end{table}
}
\end{center}

\begin{figure}
	\includegraphics[width=75mm]{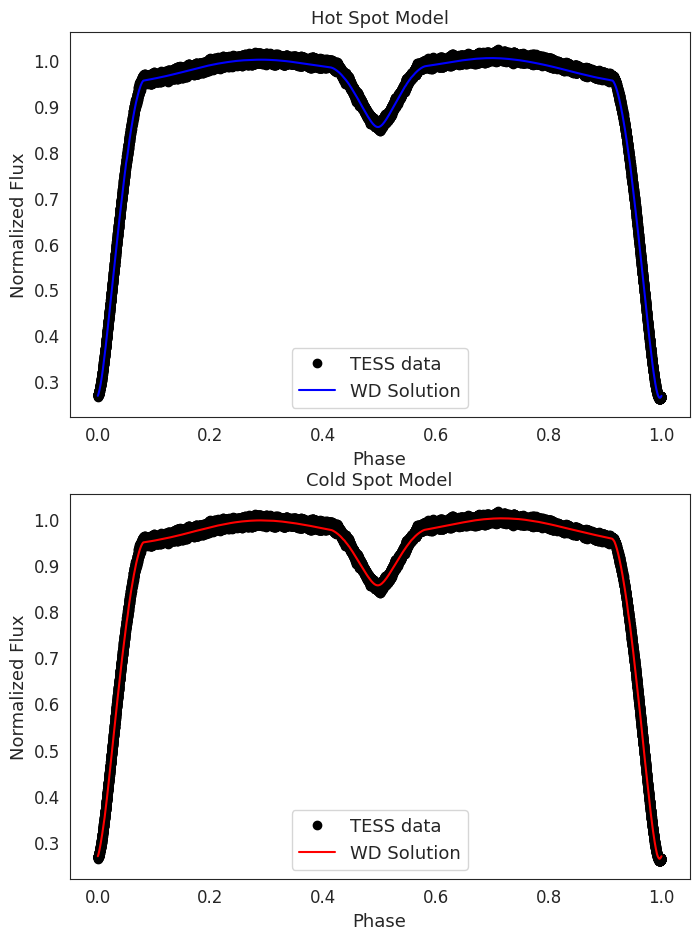}
    \caption{TESS light curves of Y\,Cam with binary models (solid lines). The upper and lower panels show the consistency of hot and cool spot models with the observation, respectively.}
    \label{fig:figure4}
\end{figure}

\section{Binary Modelling}

To determine the accurate fundamental stellar parameters of Y\,Cam, the TESS light curve was analyzed using the results of $v$$_{r}$ analysis. In the binary modeling of Y\,Cam, we utilized the Wilson\,-\,Devinney (WD) code \citep{1971ApJ...166..605W} combined with Monte-Carlo techniques \citep{2004AcA....54..299Z,2010MNRAS.408..464Z}. Given that Y\,Cam is a known Algol-type binary system \citep{2012A&A...548A..79A} we adopted a semi-detached binary configuration (mode 5) during the analysis. We initially observed that the secondary minimum of the system was affected by pulsations. Therefore, we first theoretically modeled the pulsations using the PERIOD04 program \citep{2005CoAst.146...53L} and removed them from the observational light curve before the binary modeling. All sector data were combined and used for the analysis. After removing the pulsations, we checked the maximum levels of the light curve and observed spot-like variations. consequently, we created two models by normalizing the light curve to 1 at 0.25 and 0.75 phases. For the 0.25 and 0.75 phases, we assumed a hot spot and a cold spot on the primary and secondary components, respectively. During the binary modeling, the $T_{\rm eff1}$ value of the primary star was taken to be 7800K\,$\pm$\,200 K from the spectral modeling. The logarithmic limb darkening coefficients was obtained from \citet{1993AJ....106.2096V}, based on $T_{\rm eff}$ values of the binary components. Given that the primary and secondary components have radiative and convective envelopes, respectively, we set the gravity darkening coefficients as $g_1$\,=\,1.0 and $g_2$\,=\,0.32, respectively. Additionally, the bolometric albedos were set to $A_1$\,=\,1.0 and $A_1$\,=\,0.5, for primary and secondary components. The semi-major axis ($a$) and mass ratio ($q$) values were taken from the $v$$_{r}$ curve analysis. Phase shift ($\phi$), effective temperature of secondary component ($T_{\rm eff2}$), surface potentials ($\Omega_1$, $\Omega_2$), inclination (\emph{i}) were derived from the analysis of binarity. The binary modeling is shown in Fig.~\ref{fig:figure4}. Results of light curve analysis and spot parameters are given in Table~\ref{tab:table_3}. We obtained the sum of squares\footnote{Sum of Squares: $\Sigma(O-C)^2$} as 0.0433 for the cold spot model and as 0.0294 for the hot spot model. The interaction of the gas stream with the atmosphere of the gainer can generate a high-temperature zone. The variable He I lines in the spectra of A-F type gainers serve as reliable indicators of the existence of an accretion hot spot in the atmosphere \citep{2018A&A...615A.131L}. Hence, we take into the results of the hot spot model for further analysis. Despite the $O\,-\,C$ analysis finding, a possible third component has not been detected in the binary modeling. It is essential to acknowledge that the uncertainty rate of this second $O\,-\,C$ analysis is high, primarily due to the lower sensitivity of the visual and photographic minimum times, which are more abundant, and the inherent scattering of the data.

Consequently, we calculated the fundamental stellar parameters by using the results of light curve analysis using the Kepler, Pogson equations and the bollometric corrections given by \cite{2020MNRAS.496.3887E}. In addition, we calculated the distance of the system by taking into account the E(B-V) value calculated in section 4 and another E(B-V) value of $0^m$.039\,$\pm$\,0.002 obtained from the SFD dust map module \citep{1998ApJ...500..525S}. Consequently, the distance of the system determined as 748\,$\pm$\,76\,pc and 789\,\,$\pm$\,81 pc from Bayestar19 and SFD, respectively. The distance value calculated using the SFD E(B-V) is closer to the value of 799 pc given in the Gaia DR3 catalog by \citet{2022yCat.1355....0G}. The calculated fundamental astrophysical parameters are given in Table~\ref{tab:table_4}.

\begin{center}
    \centering
    {\scriptsize
    \begin{table}
    \caption{Results of the binary modeling.}
    \label{tab:table_3}
    \begin{tabular}{p{2.3cm}p{2.4cm}p{2.4cm}}
    \hline\noalign{\vskip3pt}
    Parameter		   &Hot Spot          &Cold Spot \\
                        & Model            &  Model  \\
    \hline\noalign{\vskip3pt}
    $i$	($^{\circ}$) &85.90\,$\pm$\,0.01    & 85.59\,$\pm$\,0.02  \\
    $\phi$             &-0.0029\,$\pm$\,0.0001  & -0.0033\,$\pm$\,0.0001 \\
    $q^a$                &0.237\,$\pm$\,0.010     &0.237\,$\pm$\,0.010  \\
    $T^a_1$ (K)          &7800\,$\pm$\,200        & 7800\,$\pm$\,200      \\
    $T_2$ (K)          &4521\,$\pm$\,250        & 4368\,$\pm$\,200      \\
    $\Omega_1$         &4.303\,$\pm$\,0.004   & 4.270\,$\pm$\,0.005 \\
    $\Omega_2$        &2.324\,$\pm$\,0.006    & 2.324\,$\pm$\,0.005 \\
    $l_1/(l_1+l_2)$      &0.837\,$\pm$\,0.003 & 0.860\,$\pm$\,0.003 \\
    $l_2/(l_1+l_1)$      &0.1623\,$\pm$\,0.001 & 0.140\,$\pm$\,0.001\\
    $r_{1 (mean)}$     &0.2480\,$\pm$\,0.0002   & 0.2501 \,$\pm$\,0.0002\\
    $r_{2 (mean)}$     &0.2621\,$\pm$\,0.0002   & 0.2621\,$\pm$\,0.0003\\
    $Sum$ $of squares$           &0.0294                &0.0464                 \\
    \multicolumn{3}{c}{\textbf{Spot Parameters}}\\
    \hline\noalign{\vskip3pt}
    Parameter &  & \\
    \hline\noalign{\vskip3pt}
    $\varphi$ ($^{\circ}$) & $90.00^a$            & $90.00^a$ \\
    $\lambda$ ($^{\circ}$)   & 173.25 $\pm$ 0.14  & 314.95 $\pm$ 0.14\\
    $\theta$  ($^{\circ}$)     & 14.60 $\pm$ 0.04    & 7.06 $\pm$ 0.04 \\
    TF  & 1.103 $\pm$ 0.002 & 0.604 $\pm$ 0.006\\
    \hline\noalign{\vskip3pt}
    \end{tabular}

    \begin{tabnote}
    {\textbf{Notes:} Subscript $1$ and $2$ represent the primary and secondary components, respectively. $\varphi$ is the co-latitude of the spot, $\lambda$ is the longitude of the spot, $\theta$ is the angular radius of the spot, and TF represents the temperature factor of the spot.}
    \end{tabnote}
\end{table}
}
\end{center}

\begin{center}
    \centering
    {\scriptsize
    \begin{table*}
    \caption{Calculated fundamental stellar parameters of the binary components of Y\,Cam.}
    \label{tab:table_4}
    \begin{tabular}{p{3cm}p{3cm}p{3cm}p{3cm}p{3cm}}
    \hline\noalign{\vskip3pt}
    & Hot Spot Model &  & Cold Spot Model &  \\
    \hline\noalign{\vskip3pt}
    Parameter       & Primary           & Secondary        & Primary           & Secondary \\
    \hline\noalign{\vskip3pt}
    $M$ ($M_\odot$) &2.04\,$\pm$\,0.07  &0.48\,$\pm$\,0.03 &2.04\,$\pm$\,0.09  &0.48\,$\pm$\,0.03\\
    $R$ ($R_\odot$) &3.14\,$\pm$\,0.06  &3.33\,$\pm$\,0.06 &3.17\,$\pm$\,0.08  &3.33\,$\pm$\,0.07\\
    log L/$L_\odot$ &1.52\,$\pm$\,0.04  &0.64\,$\pm$\,0.04 &1.52\,$\pm$\,0.04  &0.59\,$\pm$\,0.04\\
    log g (cgs)     &3.75\,$\pm$\,0.01  &3.08\,$\pm$\,0.03 &3.75\,$\pm$\,0.02  &3.08\,$\pm$\,0.03 \\
    $M_{bol}$ (mag) &0.95\,$\pm$\,0.08  &3.15\,$\pm$\,0.09 &0.93\,$\pm$\,0.09  &3.28\,$\pm$\,0.11 \\
    $M_{v}$ (mag)   &0.99\,$\pm$\,0.05  &3.72\,$\pm$\,0.08 &0.97\,$\pm$\,0.07  &3.88\,$\pm$\,0.08 \\
    $V_{syn}$ (km/s)  &48\,$\pm$\,8     &51\,$\pm$\,10      &49\,$\pm$\,9      &51\,$\pm$\,10\\
    \hline\noalign{\vskip3pt}
    \end{tabular}
\end{table*}
}
\end{center}

\section{Frequency Analysis}

\begin{center}
    \centering
    {\scriptsize
\begin{table}
\caption{Results of the frequency analysis. f$_{orb}$ represent the frequency of the orbital period (0.3025\,d$^{-1}$).}
\label{tab:table_5}
\begin{tabular}{lcccc}
\hline
                & Frequency       & Amp           & SNR     & Q            \\
                & (d$^{-1}$)      & (mmag)        &         & (d)          \\ 
                & $\pm$\,0.00004  & $\pm$\,0.02   &         & $\pm$\,0.002 \\
\hline
$f_1-4f_{orb}$  &13.84035          & 0.52          & 8.2     & 0.019 \\
$f_1-2f_{orb}$  &14.44526          & 1.08          & 14.4    & 0.018 \\
$f_1$           &15.04699          & 4.69          & 20.8    & 0.017 \\
$f_2-2f_{orb}$  &18.69808          & 0.45          & 5.7     & 0.014 \\
$f_2-2f_{orb}$  &18.68976          & 0.56          & 7.1     & 0.014 \\
$f_2$           &19.31290          & 1.31          & 16.6    & 0.013 \\
$f_2+2f_{orb}$  &19.90617          & 0.57          & 8.9     & 0.013 \\
$f_3$           &14.98884          & 1.36          & 17.0    & 0.017 \\
$f_4$           &19.89150          & 0.69          & 10.6    & 0.013 \\
$f_5$           &18.95368          & 0.56          & 7.0     & 0.014 \\
$f_6$           &18.30689          & 0.48          & 6.3     & 0.014 \\
$f_6+2f_{orb}$  &17.70198          & 0.36          & 5.2     & 0.014 \\
$f_7$           &14.56605          & 0.35          & 4.5     & 0.018 \\
\hline
\end{tabular}
\end{table}
}
\end{center}

We performed a frequency analysis to investigate the pulsation structure of Y\,Cam. Utilizing the PERIOD04 software \citep{2005CoAst.146...53L} we sought individual frequencies from the observational data, along with their combinations and harmonics. Before the analysis, we removed the binarity effect from the observational light curve by incorporating a fit comprising the frequency of the orbital period and its harmonics \citep{2022MNRAS.510.1413K}. Subsequently, we carried out frequency analysis on the residual light curve, simultaneously analyzing all sector data. Our search spanned frequencies between approximately $4-80$ d$^{-1}$ with a significant limit set at 4.5 $\sigma$ SNR \citep{2021AcA....71..113B} to identify $\delta$\ Sct-type pulsations. As a result, we detected 13 significant frequencies. The findings of the frequency analysis are listed in Table~\ref{tab:table_5}. The dominant pulsating frequency was determined to be 15.047 d$^{-1}$ with an amplitude of 4.69 mmag. Additionally, for all frequencies, we calculated the pulsation constant $Q$ = $P$\(\sqrt{\rho/\rho_\odot}\) with the range of $0.013\,-\,0.019$\,d using $M$ and $R$ values of the primary $\delta$\ Sct pulsator obtained in this study. This corresponds to pulsation of the third and fifth radial overtone \citep{1981ApJ...249..218F}.

\begin{figure}
        \includegraphics[width=83mm]{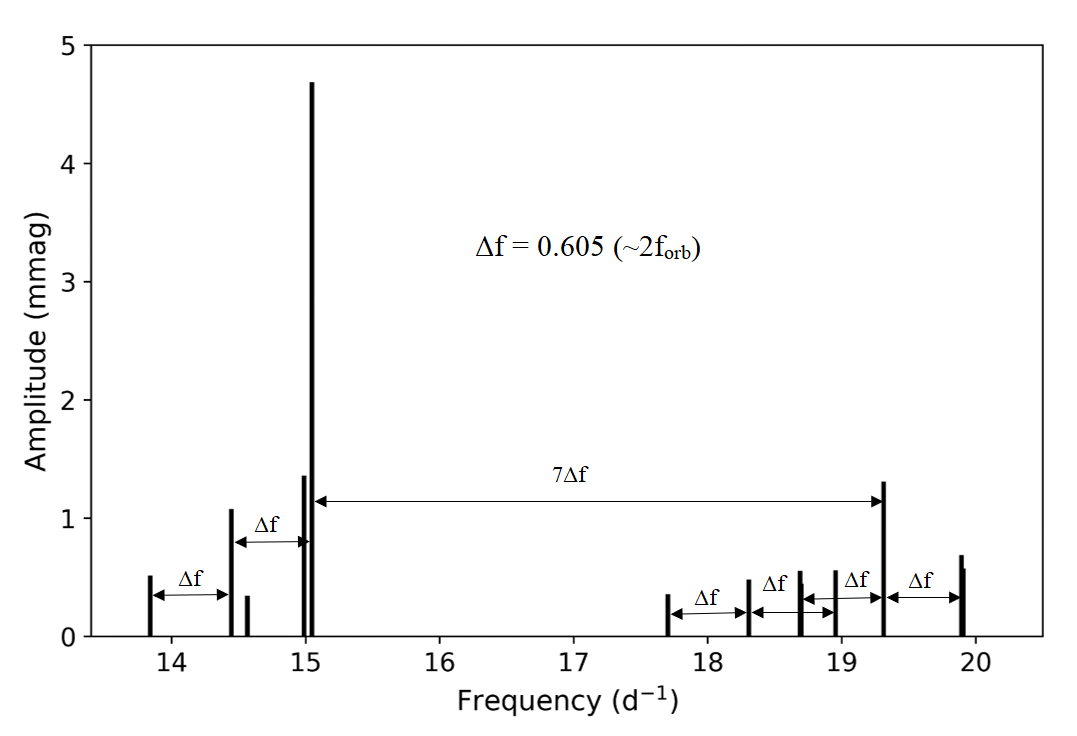}
    \caption{Frequency spectra of Y\,Cam.\,The frequencies that are over the 4.5-$\sigma$ level are represented.\,The frequency of the orbital period (f$_{orb}$) is 0.3025\,d$^{-1}$.}
    \label{fig:figure5}
\end{figure}

\begin{figure}
        \includegraphics[width=85mm]{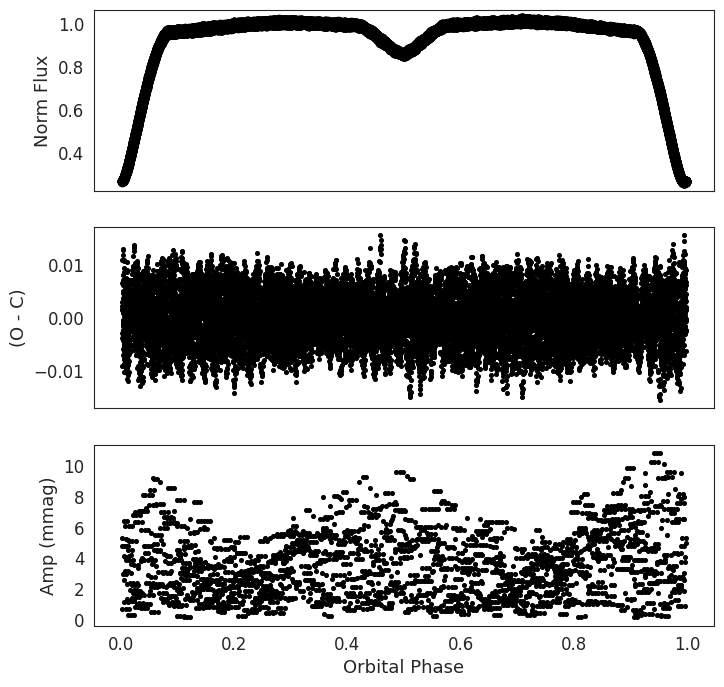}
    \caption{Upper and middle panels show the phased light curve of Y\,Cam, and its residual, respectively. The bottom panel displays the amplitude variation of the highest amplitude frequency according to the orbital phase. }
    \label{fig:figure6}
\end{figure}

We examined the obtained frequencies to ascertain whether the system exhibits characteristics of a tidally tilted pulsator. Notably, some frequencies are evenly spaced by the twice orbital period's frequency (0.605\,d$^{-1}$) around $f_1$ and $f_2$, suggesting the presence of modulated frequencies, potentially indicating Y\,Cam as a possible tidally tilted pulsator \citep{2020NatAs...4..684H,2021MNRAS.503..254R}. This frequency distribution of Y\,Cam is shown in Fig.~\ref{fig:figure5}. Furthermore, it is known that the pulsation amplitude varies throughout the orbital phase in the tidally tilted systems \citep{2020NatAs...4..684H,2021MNRAS.503..254R}. Therefore, we investigated whether the system demonstrates this type of variation using the highest amplitude frequency ($f_1$). As depicted in Fig.~\ref{fig:figure6} the amplitude reaches its maximum at around 0.5 and 1.0 phase. During the primary eclipse, when the pulsating component's surface is obscured by the secondary component, and conversely, during the secondary eclipse, when the pulsating component's surface is aligned with the line of sight, the amplitude of the frequency notably increases compared to phases 0.25 and 0.75. This observed behavior is indicative of tidally tilted zonal mode pulsations \citep{2022Galax..10...97M}. In such pulsators, notably the maximum amplitude is observed at phase 0.5, aligning with the line of sight. Additionally, the occurrence of double-peak amplitude variations around phase 1.0 suggests the visibility of the opposite pulsation pole, a phenomenon attributed to the periodic spatial filter effect \citep{2003ASPC..292..369G,2018MNRAS.475.4745M}. It's worth noting that similar amplitude modulation has been observed in some other binary systems \citep{2021PASJ...73..809L}. 


\begin{figure*}
    \includegraphics[width=\textwidth]{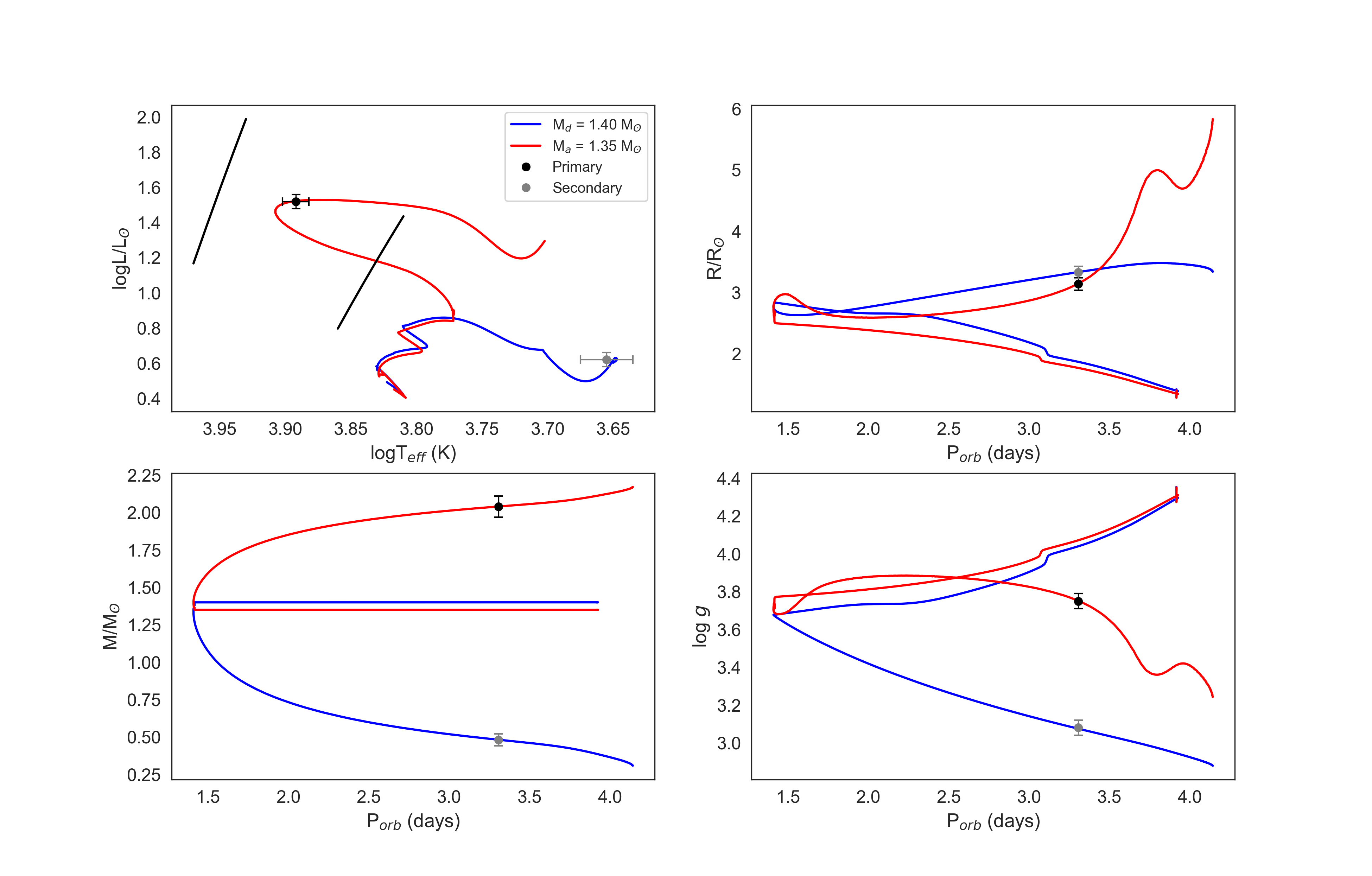}
    \caption{Top left panel shows the position of components in the Hertzsprung-Russell diagram with the $\delta$\ Sct instability strip (black lines, \cite{2019MNRAS.485.2380M}). Other panels show the compatibility of the model with observation data given $M$, $R$, and log $g$ values. Since the error bar of $P_{orb}$ smaller than size of the symbols, it is not given in the figure.}
    \label{fig:figure7}
\end{figure*}

\section{Evolutionary Modeling}

During the evolution of a classical Algol-type binary, the system undergoes several distinct stages. Initially, it starts as a detached binary system and then transitions into a semi-detached binary configuration. During the evolution between these binary configurations, in the first phase, the initially more massive binary component evolves and the system undergoes a fast mass transfer stage. Eventually, the originally less massive component gains mass and evolves into a more massive object, while the initially more massive one turns into a less massive, evolved star. Following this phase, a slow mass transfer process ensues between the binary components with Roche lobe overflow \citep{1959cbs..book.....K}. Y\,Cam should have undergone similar stages of evolution, indicating that a single-star evolution model would be inadequate to fully describe its evolutionary path.

The evolution process of Y\,Cam was investigated using Modules of Experimental Stellar Astrophysics (MESA) code with binary module \citep{2011ApJS..192....3P,2013ApJS..208....4P,2015ApJS..220...15P,2018ApJS..234...34P,2019ApJS..243...10P,2023ApJS..265...15J}. Models were generated under the assumption of non-conservative mass transfer. During the model calculations, the OPAL opacity tables were utilized \citep{2002ApJ...576.1064R}. We adopted an initial metal abundance based on the solar metal composition given by \citet{2009ARA&A..47..481A}. The overshooting parameter was set to 0.20 for both components according to \citet{2000A&A...360..952H}, and the mixing length parameter $\alpha_{MLT}$ was taken as 2.0 considering the theoretical modeling of the $\delta$ Sct instability strip \citep{2005A&A...435..927D}. Additionally, we assumed mass loss using the $\alpha$ parameter, representing the fractional mass loss from the system near the secondary component as defined by \citet{1997A&A...327..620S}. Also, magnetic braking was considered during the calculation of evolutionary modeling \citep{1983ApJ...275..713R}. Stellar rotation effects were accounted for in the models with initial rotation velocities set to 30 kms$^{-1}$ and 27 kms$^{-1}$ for donor and gainer stars, respectively, following \citet{1976asqu.book.....A}. Given that the initial mass, mass ratio, metallicity, and orbital period of binary systems influence their evolution, we thoroughly investigated the initial masses of the components, initial orbital period, and also metallicity ($Z$) values during the analysis. This exhaustive approach aimed to identify the best fitting models that accurately reflect the current state of Y\,Cam. The $Z$ parameter was explored within the range of 0.010 to 0.030. Ultimately, it was found to be $Z$\,=\,0.017\,$\pm$\,0.002 for the donor and $Z$\,=\,0.015\,$\pm$\,0.002 for the gainer star. The magnetic braking parameter was determined to be 3.585\,$\pm$\,0.002 after exploration within the range of 3.0 to 4.0. The most compatible model with observations was achieved with initial masses for donor and gainer of $M_d$\,=\,1.40\,$\pm$\,0.02\,$M_\odot$, $M_a$\,=\,1.35\,$\pm$\,0.02\,$M_\odot$ respectively, along with an initial orbital period $P_{orb}$\,=\,3.916 days. The error bars on the parameters were estimated considering the 1$\sigma$ uncertainty. The obtained evolutionary tracks of Y\,Cam are shown in Fig.~\ref{fig:figure7}.

The mass loss amount, represented by the $\alpha$ parameter, was held constant throughout both the high-mass transfer and slow-mass transfer stages. During the analysis, various $\alpha$ parameters ranging from 0.10 to 0.25 were tested. Ultimately, the most suitable model for the current situation of the Y\,Cam was identified with a $\alpha$ parameter of 0.198.
According to the model, the first Roche-lobe overflow (RLOF) commenced at $P_{orb}$\,=\,1.431\,$\pm$\,0.003 days. During the RLOF stage, the mass transfer rate has ranged between 2\,x\,10$^{-7}$\,$-$\, 9\,x\,10$^{-9}$\, M$_\odot$\,yr$^{-1}$ and the mass loss rate has changed between 10$^{-8}$\,$-$\,10$^{-10}$ M$_\odot$\,yr$^{-1}$. Eventually, the masses of the donor (current secondary) and the gainer (current primary) stars reached $M_d$\,=\,0.48\,$\pm$\,0.03\,$M_\odot$ and $M_a$\,=\,2.04\,$\pm$\,0.07\,$M_\odot$, respectively, when $P_{orb}$\,=\,3.307\,$\pm$\,0.004 days.

\begin{figure}
    \includegraphics[width=85mm]{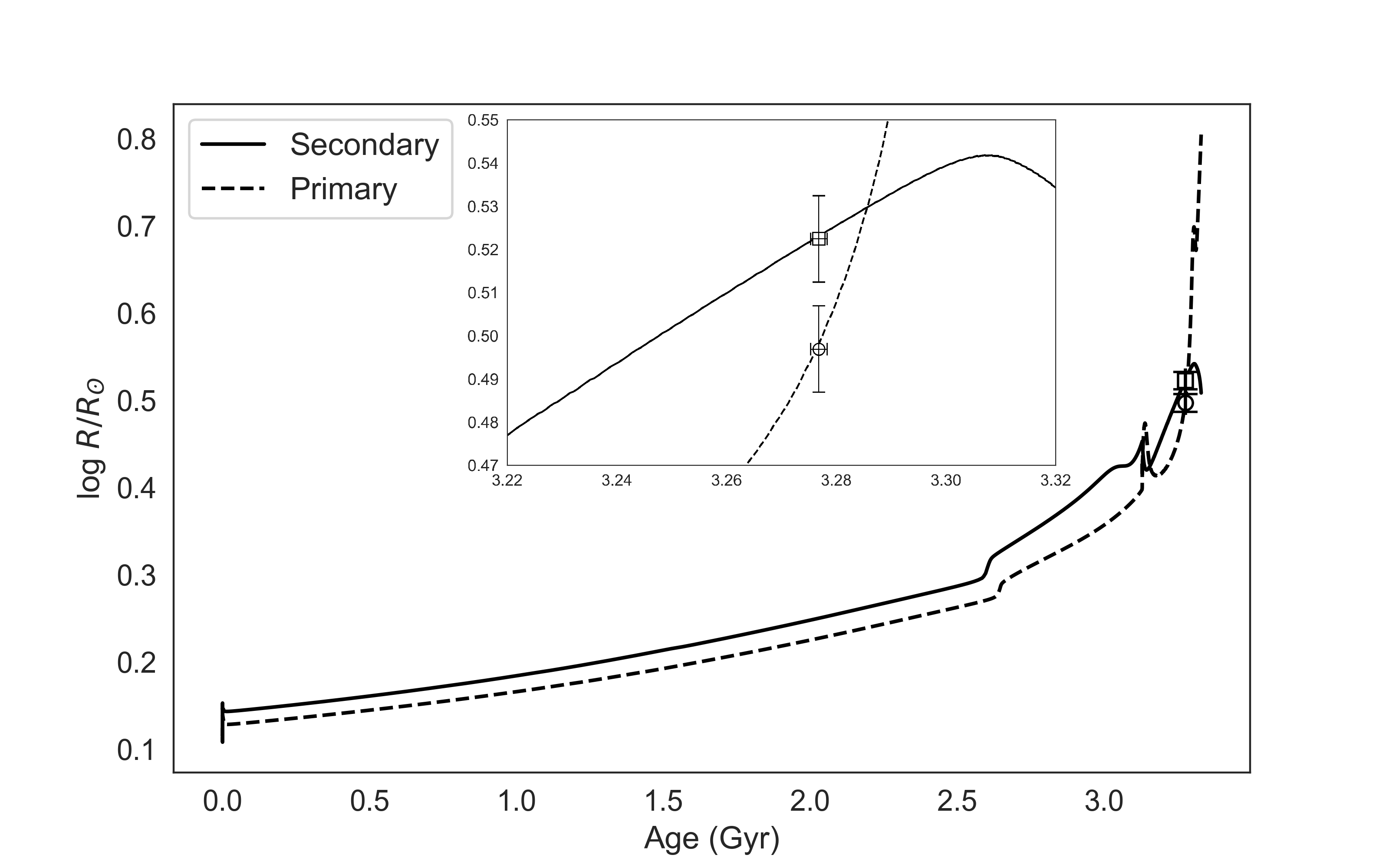}
    \caption{The position of primary (square) and secondary (circle) components of Y Cam in the Age - log $R$ diagram.}
    \label{fig:figure8}
\end{figure}

The age of the system was determined to be 3.28\,$\pm$\,0.09\,Gyr as illustrated in Fig.\,\ref{fig:figure8}, representing the point at which it attains its current binary system configuration. Towards the end of the model, the donor star evolves into a pre-He white dwarf with a mass of $M_d$\,=\,0.308\,$M_\odot$, while the gainer star evolves into the red giant stage with a mass of $M_a$\,=\,2.171\,$M_\odot$. When $P_{orb}$\,=\,4.142\,$\pm$\,0.005 days (at an age of 3.32), the system transitions into a contact binary, marking the conclusion of the model. This model effectively elucidates the current evolutionary status of the system providing insight into radii and surface gravity (log $g$) of component stars. 

\section{Discussion and Conclusion}

\begin{figure}
        \includegraphics[width=85mm]{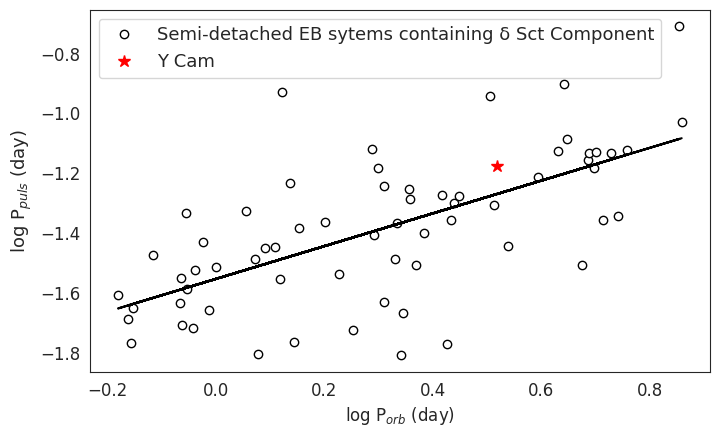}
    \caption{Position of the pulsating component of Y\,Cam (star symbol) in the $log P_{puls} - log P_{orb}$ diagram. The black line is plotted fit by using linear correlation expression \citep{2017MNRAS.470..915K}. The circles represent the known semi-detached EB systems with $\delta$ Sct components \citep{2017MNRAS.470..915K}.}
    \label{fig:figure9}
\end{figure}

\begin{figure*}

        \includegraphics[width=\textwidth]{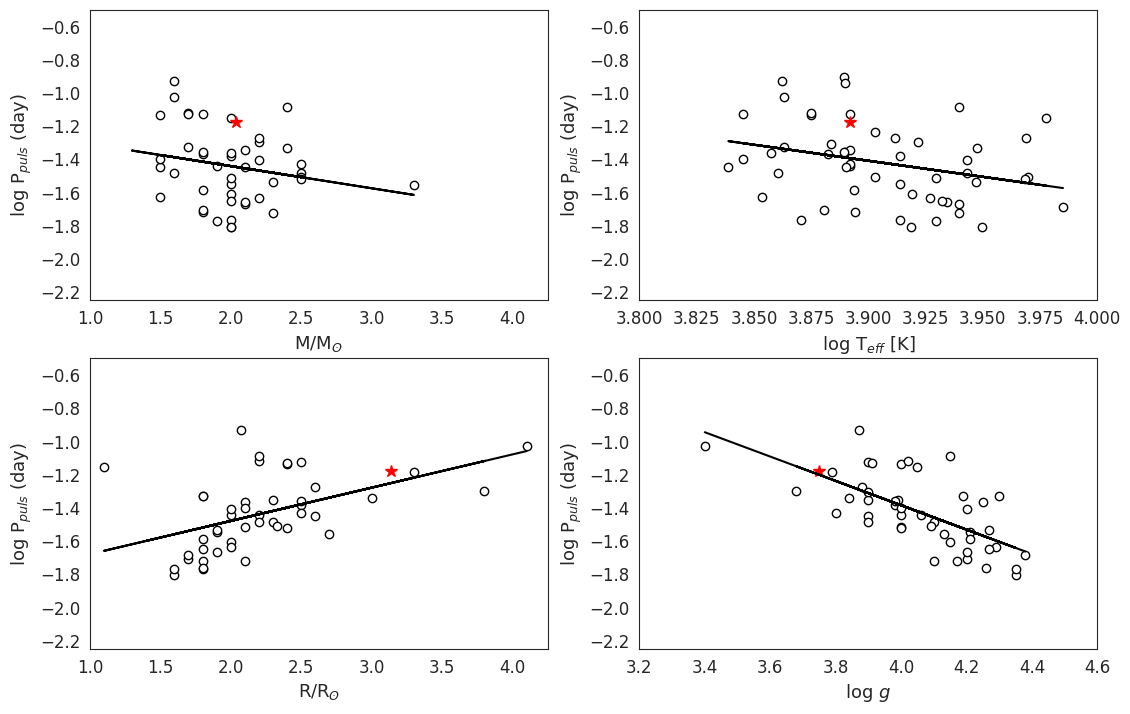}
    \caption{Upper left panel: $log P_{puls} - M/M_{\odot}$ relationship. Upper right panel: $log P_{puls} - log T_{eff}$ relationship. Bottom left panel: $log P_{puls} - R/R_{\odot}$ relationship. Bottom right panel: $log P_{puls} - log g$ relationship. Circles represent semi-detached EB systems containing $\delta$\ Sct components given by \citet{2017MNRAS.470..915K}. Black lines are the fits plotted using linear correlation expression given by \cite{2017MNRAS.470..915K} and the red star represents the pulsating component of Y\,Cam. }
    \label{fig:figure10}
\end{figure*}

In this study, we present the results of the photometric and spectral analysis of Y\,Cam, an Algol-type eclipsing binary system. Double-lined $v$$_{r}$ analysis of the system was initially conducted by \citet{2015AJ....150..131H}. In the current study, we contribute two new $v$$_{r}$ values of the primary component obtained from ELODIE spectra, thereby updating the $v$$_{r}$ analysis. Notably, the results of our analysis are consistent with those of \citet{2015AJ....150..131H} within the respective uncertainties.

Binary modeling was performed utilizing the high-quality TESS light curve following the determination of $T_{\rm eff}$ values of the component stars through the composite synthetic spectra modeling. Two different binary models were generated assuming the presence of both a hot and a cold spot. Various mechanisms have been proposed to elucidate the additional mass loss observed in Algol-type systems. Of these mechanisms, the hot spot is considered more convenient to explain mass loss. Consequently, we relied on the results obtained from the hot spot model while deriving the fundamental stellar parameters of Y\,Cam. Comparison of these parameters with those from \citet{2015AJ....150..131H} reveals close agreement within the error bars. However, slight discrepancies are noted in the $M_{bol}$ and $M_{v}$ parameters. These variations may stem from differences in the quality of the light curve and discrepancies in the methodology used for determining $T_{\rm eff}$ values.

In the continuation of the study, orbital period analysis was carried out to examine the mass transfer event. Consequently, we determined an increase in the orbital period. Also, we detected another cyclic variation in the residual after removing the parabolic variation caused by the mass transfer. According to the analysis of this variation, there may be a third object with a mass of 1.74\,$M_\odot$ connected to the system. Although it is a fairly high mass, we did not find the light contribution belonging to the third body during the binary modeling. One should keep in mind that since the O-C variation mostly consists of visual and photographic minimum times, being sure about the sinusoidal variation related to the third body more reliable data is needed. 

To determine the pulsation feature of Y Cam, we carried out a frequency analysis. Three sectors of TESS SC data were used during the analysis. Eventually, the dominant pulsation frequency was found to be $F_1 = 15.05$ d$^{-1}$ and the pulsating amplitude is $A = 4.65$ mmag. This frequency seems to be consistent with \citet{2010MNRAS.408.2149R}. For this frequency, the pulsating constant was calculated as $Q = 0.017$ d, and this value corresponds to the third radial overtone \citep{1981ApJ...249..218F}. In addition, we examined whether the system is a tidally tilted pulsator. The amplitude variation of the system and the frequency modulation with the orbital phase were examined. As a result, we concluded that the system is a candidate of tidally tilted pulsators.

The position of the pulsating component of Y\,Cam on the well-known relationship between pulsation period ($P_{puls}$) and $P_{orb}$ was examined in Fig.~\ref{fig:figure9}. Its position is compatible with the correlation given for semi-detached EB systems containing $\delta$\ Sct component \citep{2017MNRAS.470..915K}. Also, the position of the pulsating component is shown in different known relationships between $P_{puls}$ and other parameters shown in Fig.~\ref{fig:figure10}. Among these, the relationships with log $g$ and $R$ are notably consistent with given correlations. It is worth noting that the utilized relationships were sourced from \citet{2017MNRAS.470..915K}. However, it is important to acknowledge that these relationships may lack sensitivity, as most of the parameters of these objects were not determined based on radial velocity, spectral analysis, and high-quality space data.

The evolutionary status of Y\,Cam was further explored through modeling using the determined fundamental stellar parameters and MESA binary evolutionary models. This analysis allowed us to ascertain the initial masses, the initial rotational velocity of the binary components, and the initial $P_{orb}$ value of the system. The derived Z values for the primary and secondary binary components were found to be 0.017\,$\pm$\,0.002 and 0.015\,$\pm$\,0.002, respectively. It is known that binary stars are formed in the same circumstellar area, suggesting they should have similar Z values. In the case of Y\,Cam, the binary components have similar Z abundances within the error bars which is suitable for the expected scenario.  

The comprehensive analysis of Y\,Cam marks the first detailed investigation into the system's evolutionary status, utilizing binary evolutionary models that account for mass transfer, rotation, and magnetic braking effects. Studying such systems is crucial for understanding the impact of mass transfer on evolution and pulsation dynamics within binary systems.

\section*{Acknowledgements}

We thank Dr. Mkrtichian for his helpful comments that have improved the manuscript. This article is a part of the master thesis of E\c{C}. This study has been partially supported by the Scientific and Technological  Research  Council (TUBITAK) project through 120F330. The authors thank Dr. F. Alicavus for his valuable comments on the evolutionary modeling analysis. The TESS data presented in this paper were obtained from the Mikulski Archive for Space Telescopes (MAST). Funding for the TESS mission is provided by the NASA Explorer Program. This work has made use of data from the European Space Agency (ESA) mission Gaia (http://www.cosmos.esa.int/gaia), processed by the Gaia Data Processing and Analysis Consortium (DPAC, http://www.cosmos.esa.int/web/gaia/dpac/consortium). Funding for the DPAC has been provided by national institutions, in particular, the institutions participating in the Gaia Multilateral Agreement. This research has made use of the SIMBAD database, operated at CDS, Strasbourg, France. 



\begin{thebibliography}{80}

\bibitem[Aerts et al.(2010)]{2010aste.book.....A} Aerts, C., Christensen-Dalsgaard, J., \& Kurtz, D.~W.\ 2010, Asteroseismology, Astronomy and Astrophysics Library. ISBN 978-1-4020-5178-4. Springer Science+Business Media B.V., 2010, p.. doi:10.1007/978-1-4020-5803-5

\bibitem[Alfonso-Garz{\'o}n et al.(2012)]{2012A&A...548A..79A} Alfonso-Garz{\'o}n, J., Domingo, A., Mas-Hesse, J.~M., et al.\ 2012, \aap, 548, A79. doi:10.1051/0004-6361/201220095

\bibitem[Allen(1976)]{1976asqu.book.....A} Allen, C.~W.\ 1976.\ Astrophysical Quantities.\ Astrophysical Quantities, London: Athlone (3rd edition), 1976.

\bibitem[Alonso et al.(1996)]{1996A&AS..117..227A} Alonso, A., Arribas, S., Martinez-Roger, C.\ 1996.\ Determination of effective temperatures for an extended sample of dwarfs and subdwarfs (F0-K5)..\ Astronomy and Astrophysics Supplement Series 117, 227–254.

\bibitem[Asplund et al.(2009)]{2009ARA&A..47..481A} Asplund, M., Grevesse, N., Sauval, A.~J., et al.\ 2009, \araa, 47, 481. doi:10.1146/annurev.astro.46.060407.145222

\bibitem[Applegate(1992)]{1992ApJ...385..621A} Applegate, J.~H.\ 1992.\ A Mechanism for Orbital Period Modulation in Close Binaries.\ The Astrophysical Journal 385, 621. doi:10.1086/170967

\bibitem[Baran and Koen(2021)]{2021AcA....71..113B} Baran, A.~S., Koen, C.\ 2021.\ A Detection Threshold in the Amplitude Spectra Calculated from TESS Time-Series Data.\ Acta Astronomica 71, 113–121. doi:10.32023/0001-5237/71.2.3

\bibitem[Broglia(1973)]{1973IBVS..823....1B} Broglia, P.\ 1973, Information Bulletin on Variable Stars, 823, 1

\bibitem[Broglia \& Marin(1974)]{1974A&A....34...89B} Broglia, P. \& Marin, F.\ 1974, \aap, 34, 89


\bibitem[Catanzaro(2014)]{2014dapb.book...97C} Catanzaro, G.\ 2014, Determination of Atmospheric Parameters of B, 97. doi:10.1007/978-3-319-06956-2\_9

\bibitem[Ceraski(1903)]{1903AN....162..205C} Ceraski, W.\ 1903, Astronomische Nachrichten, 162, 205. doi:10.1002/asna.19031621307

\bibitem[Chen et al.(2022)]{2022ApJS..263...34C} Chen, X. and 8 colleagues 2022.\ Detection of {\ensuremath{\delta}} Scuti Pulsators in the Eclipsing Binaries Observed by TESS.\ The Astrophysical Journal Supplement Series 263. doi:10.3847/1538-4365/aca284

\bibitem[Cutri et al.(2003)]{2003yCat.2246....0C} Cutri, R.~M. and 24 colleagues 2003.\ VizieR Online Data Catalog: 2MASS All-Sky Catalog of Point Sources (Cutri+ 2003).\ VizieR Online Data Catalog.


\bibitem[Dupret et al.(2005)]{2005A&A...435..927D} Dupret, M.-A., Grigahc{\`e}ne, A., Garrido, R., Gabriel, M., Scuflaire, R.\ 2005.\ Convection-pulsation coupling. II. Excitation and stabilization mechanisms in {\ensuremath{\delta}} Sct and {\ensuremath{\gamma}} Dor stars.\ Astronomy and Astrophysics 435, 927–939. doi:10.1051/0004-6361:20041817

\bibitem[Erdem and {\"O}zt{\"u}rk(2014)]{2014MNRAS.441.1166E} Erdem, A., {\"O}zt{\"u}rk, O.\ 2014.\ Non-conservative mass transfers in Algols.\ Monthly Notices of the Royal Astronomical Society 441, 1166–1176. doi:10.1093/mnras/stu630

\bibitem[Eker et al.(2020)]{2020MNRAS.496.3887E} Eker, Z. and 8 colleagues 2020.\ Empirical bolometric correction coefficients for nearby main-sequence stars in the Gaia era.\ Monthly Notices of the Royal Astronomical Society 496, 3887–3905. doi:10.1093/mnras/staa1659

\bibitem[Fitch(1981)]{1981ApJ...249..218F} Fitch, W.~S.\ 1981.\ L=0, 1, 2, and 3 pulsation constants for evolutionary models of del SCT stars..\ The Astrophysical Journal 249, 218–227. doi:10.1086/159278

\bibitem[Gaia Collaboration(2022)]{2022yCat.1355....0G} Gaia Collaboration\ 2022.\ VizieR Online Data Catalog: Gaia DR3 Part 1. Main source (Gaia Collaboration, 2022).\ VizieR Online Data Catalog. doi:10.26093/cds/vizier.1355

\bibitem[Gamarova et al.(2003)]{2003ASPC..292..369G} Gamarova, A.~Y., Mkrtichian, D.~E., Rodriguez, E., Costa, V., Lopez-Gonzalez, M.~J.\ 2003.\ Application of the Spatial Filtration Method to RZ Cas.\ Interplay of Periodic, Cyclic and Stochastic Variability in Selected Areas of the H-R Diagram 292, 369.

\bibitem[Gray \& Corbally(1994)]{1994AJ....107..742G} Gray, R.~O. \& Corbally, C.~J.\ 1994, \aj, 107, 742. doi:10.1086/116893

\bibitem[Green et al.(2019)]{2019ApJ...887...93G} Green, G.~M., Schlafly, E., Zucker, C., et al.\ 2019, \apj, 887, 93. doi:10.3847/1538-4357/ab5362

\bibitem[Handler et al.(2020)]{2020NatAs...4..684H} Handler, G. and 14 colleagues 2020.\ Tidally trapped pulsations in a close binary star system discovered by TESS.\ Nature Astronomy 4, 684–689. doi:10.1038/s41550-020-1035-1

\bibitem[Herwig(2000)]{2000A&A...360..952H} Herwig, F.\ 2000.\ The evolution of AGB stars with convective overshoot.\ Astronomy and Astrophysics 360, 952–968. doi:10.48550/arXiv.astro-ph/0007139

\bibitem[Hong et al.(2015)]{2015AJ....150..131H} Hong, K., Lee, J.~W., Kim, S.-L., et al.\ 2015, \aj, 150, 131. doi:10.1088/0004-6256/150/4/131

\bibitem[H{\o}g et al.(2000)]{2000A&A...355L..27H} H{\o}g, E. and 8 colleagues 2000.\ The Tycho-2 catalogue of the 2.5 million brightest stars.\ Astronomy and Astrophysics 355, L27–L30.

\bibitem[Iglesias-Marzoa et al.(2015)]{2015PASP..127..567I} Iglesias-Marzoa, R., L{\'o}pez-Morales, M., \& Jes{\'u}s Ar{\'e}valo Morales, M.\ 2015, \pasp, 127, 567. doi:10.1086/682056

\bibitem[Irwin(1959)]{1959AJ.....64..149I} Irwin, J.~B.\ 1959.\ Standard light-time curves.\ The Astronomical Journal 64, 149. doi:10.1086/107913


\bibitem[Jermyn et al.(2023)]{2023ApJS..265...15J} Jermyn, A.~S. and 21 colleagues 2023.\ Modules for Experiments in Stellar Astrophysics (MESA): Time-dependent Convection, Energy Conservation, Automatic Differentiation, and Infrastructure.\ The Astrophysical Journal Supplement Series 265. doi:10.3847/1538-4365/acae8d

\bibitem[Kahraman Ali{\c{c}}avu{\c{s}} et al.(2017)]{2017MNRAS.470..915K} Kahraman Ali{\c{c}}avu{\c{s}}, F., Soydugan, E., Smalley, B., et al.\ 2017, \mnras, 470, 915. doi:10.1093/mnras/stx1241

\bibitem[Kahraman Ali{\c{c}}avu{\c{s}} et al.(2022)]{2022MNRAS.510.1413K} Kahraman Ali{\c{c}}avu{\c{s}}, F. and 8 colleagues 2022.\ Mass transfer and tidally tilted pulsation in the Algol-type system TZ Dra.\ Monthly Notices of the Royal Astronomical Society 510, 1413–1424. doi:10.1093/mnras/stab3515

\bibitem[Kahraman Ali{\c{c}}avu{\c{s}} et al.(2022)]{2022RAA....22h5003K} Kahraman Ali{\c{c}}avu{\c{s}}, F., G{\"u}m{\"u}{\c{s}}, D., K{\i}rm{\i}z{\i}ta{\c{s}}, {\"O}., et al.\ 2022, Research in Astronomy and Astrophysics, 22, 085003. doi:10.1088/1674-4527/ac71a4

\bibitem[Kahraman Ali{\c{c}}avu{\c{s}} et al.(2023)]{2023MNRAS.524..619K} Kahraman Ali{\c{c}}avu{\c{s}}, F., {\c{C}}oban, {\c{C}}. G., {\c{C}}elik, E., Dogan, D.~S., Ekinci, O., Ali{\c{c}}avu{\c{s}}, F.\ 2023.\ Discovery of delta Scuti variables in eclipsing binary systems II. Southern TESS field search.\ Monthly Notices of the Royal Astronomical Society 524, 619–630. doi:10.1093/mnras/stad1898

\bibitem[Khaliullin \& Khaliullina(2013)]{2013ARep...57..517K} Khaliullin, K.~F. \& Khaliullina, A.~I.\ 2013, Astronomy Reports, 57, 517. doi:10.1134/S1063772913060024

\bibitem[Kim et al.(2002)]{2002A&A...391..213K} Kim, S.-L., Lee, J.~W., Youn, J.-H., et al.\ 2002, \aap, 391, 213. doi:10.1051/0004-6361:20020777

\bibitem[Kim(2023)]{2023ApJ...948...16K} Kim, S.-L.\ 2023, \apj, 948, 16. doi:10.3847/1538-4357/acc066

\bibitem[Kopal(1959)]{1959cbs..book.....K} Kopal, Z.\ 1959.\ Close binary systems.\ The International Astrophysics Series, London: Chapman \& Hall, 1959.

\bibitem[Kreiner(2004)]{2004AcA....54..207K} Kreiner, J.~M.\ 2004.\ Up-to-Date Linear Elements of Eclipsing Binaries.\ Acta Astronomica 54, 207–210.

\bibitem[Kurucz(1993)]{1993KurCD..13.....K} Kurucz, R.\ 1993, ATLAS9 Stellar Atmosphere Programs and 2 km/s grid. Kurucz CD-ROM No. 13. Cambridge, 13

\bibitem[Kwee and van Woerden(1956)]{1956BAN....12..327K} Kwee, K.~K., van Woerden, H.\ 1956.\ A method for computing accurately the epoch of minimum of an eclipsing variable.\ Bulletin of the Astronomical Institutes of the Netherlands 12, 327.

\bibitem[Lampens(2021)]{2021Galax...9...28L} Lampens, P.\ 2021, Galaxies, 9, 28. doi:10.3390/galaxies9020028

\bibitem[Lee(2021)]{2021PASJ...73..809L} Lee, J.~W.\ 2021.\ Tidally perturbed oblique pulsations in the hierarchical triple system V1031 Orionis.\ Publications of the Astronomical Society of Japan 73, 809–816. doi:10.1093/pasj/psab044

\bibitem[Lehmann et al.(2020)]{2020A&A...644A.121L} Lehmann, H., Dervi{\c{s}}o{\u{g}}lu, A., Mkrtichian, D.~E., et al.\ 2020, \aap, 644, A121. doi:10.1051/0004-6361/202039355

\bibitem[Lehmann et al.(2018)]{2018A&A...615A.131L} Lehmann, H., Tsymbal, V., Pertermann, F., Tkachenko, A., Mkrtichian, D.~E., A-thano, N.\ 2018.\ Spectroscopic time-series analysis of R Canis Majoris.\ Astronomy and Astrophysics 615. doi:10.1051/0004-6361/201629914

\bibitem[Lenz \& Breger(2005)]{2005CoAst.146...53L} Lenz, P. \& Breger, M.\ 2005, Communications in Asteroseismology, 146, 53. doi:10.1553/cia146s53

\bibitem[Liakos \& Niarchos(2017)]{2017MNRAS.465.1181L} Liakos, A. \& Niarchos, P.\ 2017, \mnras, 465, 1181. doi:10.1093/mnras/stw2756

\bibitem[Lovekin \& Deupree(2008)]{2008ApJ...679.1499L} Lovekin, C.~C. \& Deupree, R.~G.\ 2008, \apj, 679, 1499. doi:10.1086/587615

\bibitem[Masana et al.(2006)]{2006A&A...450..735M} Masana, E., Jordi, C., \& Ribas, I.\ 2006, \aap, 450, 735. doi:10.1051/0004-6361:20054021

\bibitem[Miszuda et al.(2022)]{2022MNRAS.514..622M} Miszuda, A., Ko{\l}aczek-Szyma{\'n}ski, P.~A., Szewczuk, W., Daszy{\'n}ska-Daszkiewicz, J.\ 2022.\ The eclipsing binary systems with {\ensuremath{\delta}} Scuti component - II. AB Cas.\ Monthly Notices of the Royal Astronomical Society 514, 622–639. doi:10.1093/mnras/stac1197

\bibitem[Mkrtichian et al.(2002)]{2002ASPC..259...96M} Mkrtichian, D.~E., Kusakin, A.~V., Gamarova, A.~Y., et al.\ 2002, IAU Colloq. 185: Radial and Nonradial Pulsations as Probes of Stellar Physics, 259, 96

\bibitem[Mkrtichian et al.(2004)]{2004A&A...419.1015M} Mkrtichian, D.~E., Kusakin, A.~V., Rodriguez, E., et al.\ 2004, \aap, 419, 1015. doi:10.1051/0004-6361:20040095


\bibitem[Mkrtichian et al.(2018)]{2018MNRAS.475.4745M} Mkrtichian, D.~E., Lehmann, H., Rodr{\'\i}guez, E., et al.\ 2018, \mnras, 475, 4745. doi:10.1093/mnras/stx2841

\bibitem[Mkrtichian et al.(2022)]{2022Galax..10...97M} Mkrtichian, D., Gunsriviwat, K., Lehmann, H., et al.\ 2022, Galaxies, 10, 97. doi:10.3390/galaxies10050097

\bibitem[Moultaka et al.(2004)]{2004PASP..116..693M} Moultaka, J., Ilovaisky, S.~A., Prugniel, P., et al.\ 2004, \pasp, 116, 693. doi:10.1086/422177

\bibitem[Murphy et al.(2019)]{2019MNRAS.485.2380M} Murphy, S.~J., Hey, D., Van Reeth, T., Bedding, T.~R.\ 2019.\ Gaia-derived luminosities of Kepler A/F stars and the pulsator fraction across the {\ensuremath{\delta}} Scuti instability strip.\ Monthly Notices of the Royal Astronomical Society 485, 2380–2400. doi:10.1093/mnras/stz590

\bibitem[Ouazzani et al.(2017)]{2017MNRAS.465.2294O} Ouazzani, R.-M., Salmon, S.~J.~A.~J., Antoci, V., et al.\ 2017, \mnras, 465, 2294. doi:10.1093/mnras/stw2717

\bibitem[Park et al.(2020)]{2020AJ....160..247P} Park, J.-H., Lee, J.~W., Hong, K., Koo, J.-R., Kim, C.-H.\ 2020.\ Physical Nature of the Eclipsing {\ensuremath{\delta}} Scuti Star AO Serpentis.\ The Astronomical Journal 160. doi:10.3847/1538-3881/abbef4

\bibitem[Paxton et al.(2011)]{2011ApJS..192....3P} Paxton, B., Bildsten, L., Dotter, A., Herwig, F., Lesaffre, P., Timmes, F.\ 2011.\ Modules for Experiments in Stellar Astrophysics (MESA).\ The Astrophysical Journal Supplement Series 192. doi:10.1088/0067-0049/192/1/3

\bibitem[Paxton et al.(2013)]{2013ApJS..208....4P} Paxton, B. and 10 colleagues 2013.\ Modules for Experiments in Stellar Astrophysics (MESA): Planets, Oscillations, Rotation, and Massive Stars.\ The Astrophysical Journal Supplement Series 208. doi:10.1088/0067-0049/208/1/4

\bibitem[Paxton et al.(2015)]{2015ApJS..220...15P} Paxton, B. and 12 colleagues 2015.\ Modules for Experiments in Stellar Astrophysics (MESA): Binaries, Pulsations, and Explosions.\ The Astrophysical Journal Supplement Series 220. doi:10.1088/0067-0049/220/1/15

\bibitem[Paxton et al.(2018)]{2018ApJS..234...34P} Paxton, B. and 12 colleagues 2018.\ Modules for Experiments in Stellar Astrophysics (MESA): Convective Boundaries, Element Diffusion, and Massive Star Explosions.\ The Astrophysical Journal Supplement Series 234. doi:10.3847/1538-4365/aaa5a8

\bibitem[Paxton et al.(2019)]{2019ApJS..243...10P} Paxton, B. and 16 colleagues 2019.\ Modules for Experiments in Stellar Astrophysics (MESA): Pulsating Variable Stars, Rotation, Convective Boundaries, and Energy Conservation.\ The Astrophysical Journal Supplement Series 243. doi:10.3847/1538-4365/ab2241

\bibitem[Rappaport et al.(1983)]{1983ApJ...275..713R} Rappaport, S., Verbunt, F., \& Joss, P.~C.\ 1983, \apj, 275, 713. doi:10.1086/161569

\bibitem[Rappaport et al.(2021)]{2021MNRAS.503..254R} Rappaport, S.~A. and 15 colleagues 2021.\ A tidally tilted sectoral dipole pulsation mode in the eclipsing binary TIC 63328020.\ Monthly Notices of the Royal Astronomical Society 503, 254–269. doi:10.1093/mnras/stab336

\bibitem[Ricker et al.(2015)]{2015JATIS...1a4003R} Ricker, G.~R., Winn, J.~N., Vanderspek, R., et al.\ 2015, Journal of Astronomical Telescopes, Instruments, and Systems, 1, 014003. doi:10.1117/1.JATIS.1.1.014003

\bibitem[Rogers \& Nayfonov(2002)]{2002ApJ...576.1064R} Rogers, F.~J. \& Nayfonov, A.\ 2002, \apj, 576, 1064. doi:10.1086/341894

\bibitem[Rodr{\'\i}guez \& Breger(2001)]{2001A&A...366..178R} Rodr{\'\i}guez, E. \& Breger, M.\ 2001, \aap, 366, 178. doi:10.1051/0004-6361:20000205

\bibitem[Rodr{\'\i}guez et al.(2010)]{2010MNRAS.408.2149R} Rodr{\'\i}guez, E., Garc{\'\i}a, J.~M., Costa, V., et al.\ 2010, \mnras, 408, 2149. doi:10.1111/j.1365-2966.2010.17055.x

\bibitem[Schlegel et al.(1998)]{1998ApJ...500..525S} Schlegel, D.~J., Finkbeiner, D.~P., \& Davis, M.\ 1998, \apj, 500, 525. doi:10.1086/305772

\bibitem[Sekiguchi \& Fukugita(2000)]{2000AJ....120.1072S} Sekiguchi, M. \& Fukugita, M.\ 2000, \aj, 120, 1072. doi:10.1086/301490


\bibitem[Shi et al.(2022)]{2022ApJS..259...50S} Shi, X.-. dong ., Qian, S.-. bang ., Li, L.-J.\ 2022.\ New Pulsating Stars Detected in EA-type Eclipsing-binary Systems Based on TESS Data.\ The Astrophysical Journal Supplement Series 259. doi:10.3847/1538-4365/ac59b9

\bibitem[Soberman et al.(1997)]{1997A&A...327..620S} Soberman, G.~E., Phinney, E.~S., \& van den Heuvel, E.~P.~J.\ 1997, \aap, 327, 620. doi:10.48550/arXiv.astro-ph/9703016

\bibitem[Southworth et al.(2005)]{2005A&A...429..645S} Southworth, J., Maxted, P.~F.~L., \& Smalley, B.\ 2005, \aap, 429, 645. doi:10.1051/0004-6361:20041867

\bibitem[Southworth(2013)]{2013A&A...557A.119S} Southworth, J.\ 2013, \aap, 557, A119. doi:10.1051/0004-6361/201322195

\bibitem[Southworth(2021)]{2021Univ....7..369S} Southworth, J.\ 2021, Universe, 7, 369. doi:10.3390/universe7100369

\bibitem[Stassun et al.(2019)]{2019AJ....158..138S} Stassun, K.~G. and 36 colleagues 2019.\ The Revised TESS Input Catalog and Candidate Target List.\ The Astronomical Journal 158. doi:10.3847/1538-3881/ab3467

\bibitem[Tempesti(1971)]{1971IBVS..596....1T} Tempesti, P.\ 1971, Information Bulletin on Variable Stars, 596, 1

\bibitem[Tody(1986)]{1986SPIE..627..733T} Tody, D.\ 1986, \procspie, 627, 733. doi:10.1117/12.968154

\bibitem[Torres et al.(2010)]{2010A&ARv..18...67T} Torres, G., Andersen, J., \& Gim{\'e}nez, A.\ 2010, \aapr, 18, 67. doi:10.1007/s00159-009-0025-1

\bibitem[van Hamme(1993)]{1993AJ....106.2096V} van Hamme, W.\ 1993, \aj, 106, 2096. doi:10.1086/116788

\bibitem[Wagg et al.(2024)]{2024arXiv240305627W} Wagg, T., Johnston, C., Bellinger, E.~P., et al.\ 2024, arXiv:2403.05627. doi:10.48550/arXiv.2403.05627

\bibitem[Wilson \& Devinney(1971)]{1971ApJ...166..605W} Wilson, R.~E. \& Devinney, E.~J.\ 1971, \apj, 166, 605. doi:10.1086/150986


\bibitem[Zhang et al.(2023)]{2023arXiv230516721Z} Zhang, Q.-S., Yan, L., Tao, W., et al.\ 2023, arXiv:2305.16721. doi:10.48550/arXiv.2305.16721


\bibitem[Zola et al.(2004)]{2004AcA....54..299Z} Zola, S. and 9 colleagues 2004.\ Physical Parameters of Components in Close Binary Systems: III.\ Acta Astronomica 54, 299–312.

\bibitem[Zola et al.(2010)]{2010MNRAS.408..464Z} Zola, S. and 6 colleagues 2010.\ Physical parameters of components in close binary systems - VII.\ Monthly Notices of the Royal Astronomical Society 408, 464–474. doi:10.1111/j.1365-2966.2010.17129.x

\end{thebibliography}
\end{document}